\documentclass[trackchanges,resetfootnote,twocolumn]{aastex701}
\received{June 2, 2025}
\revised{October 1, 2025}
\accepted{October 2, 2025}

\usepackage{amsmath}

\begin{document}

\title{Investigating the System Configuration of Kepler-451 through Orbital Period Variations: Dynamical and Magnetic Interpretations}

\author[0000-0003-1339-3045]{Huseyin Er}
\affiliation{Atatürk University, Faculty of Science, Department of Astronomy and Space Science, 25240, Yakutiye, Erzurum, T\"urkiye}
\email[show]{huseyin.er@atauni.edu.tr} 

\author[0000-0003-1399-5804]{Aykut Ozdonmez}
\affiliation{Atatürk University, Faculty of Science, Department of Astronomy and Space Science, 25240, Yakutiye, Erzurum, T\"urkiye}
\email{aykut.ozdonmez@atauni.edu.tr} 

\author[0000-0002-2847-8124]{M. Emir Kenger}
\affiliation{Atatürk University, Graduate School of Natural and Applied Sciences, Department of Astronomy and Astrophysics, 25240, Yakutiye, Erzurum, T\"urkiye}
\email{emirkngr@gmail.com} 

\author[0000-0003-4120-0562]{B. Batuhan Gurbulak}
\affiliation{Atatürk University, Graduate School of Natural and Applied Sciences, Department of Astronomy and Astrophysics, 25240, Yakutiye, Erzurum, T\"urkiye}
\email{burakerz@hotmail.com} 

\author[0000-0001-8131-4455]{Ilham Nasiroglu}
\affiliation{Atatürk University, Faculty of Science, Department of Astronomy and Space Science, 25240, Yakutiye, Erzurum, T\"urkiye}
\email{inasir@atauni.edu.tr} 

\author[0009-0003-2863-5577]{Murat Tekkesinoglu}
\affiliation{Atatürk University, Graduate School of Natural and Applied Sciences, Department of Astronomy and Astrophysics, 25240, Yakutiye, Erzurum, T\"urkiye}
\email{murattekkesinoglu@gmail.com} 

\author[0000-0001-5778-5679]{Ergun Ege}
\affiliation{Istanbul University, Faculty of Science, Department of Astronomy and Space Sciences, 34116, Beyazit, Istanbul, T\"urkiye}
\email{ergunege@istanbul.edu.tr} 

\author[0000-0003-3910-2285]{Nazlı Karaman}
\affiliation{Adiyaman University, Gölbaşı Vocational School, Department of Electricity and Energy, 02500, Gölbaşı, Adiyaman, T\"urkiye}
\affiliation{Adiyaman University, Astrophysics Application and Research Center, 02040, Adiyaman, T\"urkiye}
\email{nkaraman@adiyaman.edu.tr} 

\begin{abstract}
We present an analysis of eclipse timing variations in Kepler-451 using data spanning 2004–2024 from both ground- and space-based observations. Using two datasets, DS-A and DS-B, we constructed updated $O-C$ diagrams. By modeling both datasets with various LTT configurations, we tested for the presence of circumbinary companions. For DS-A, the two-companion model (LTT34) provides the best fit with RMS = 3.23 s and $\chi^2_\nu = 1.23$, while inclusion a fifth body (LTT345) does not improve the fit (RMS = 3.24 s, $\chi^2_\nu = 1.28$). For DS-B, the three-companion model (LTT345) yields the best fit (RMS = 2.31 s, $\chi^2_\nu = 1.01$), although the semi-amplitude of the inner companion (1.34 s) is smaller than the systematic error (1.81 s), suggesting that it may originate from observational or calibration systematics. Applegate-mechanism tests indicate that most signals exceed the available energy budget, while the outer LTT terms in both datasets remain consistent with the Standard model and may have a magnetic origin. Removing these magnetic terms yields dynamically stable configurations for at least $10^7$ yr. These findings support the presence of a second-generation circumbinary planet at $\sim3.4$ AU around Kepler-451, while the origin of the remaining LTT signals remains uncertain.
\end{abstract}

\keywords{Eclipsing binary minima timing method (443); Timing variation methods (1703)}


\section{Introduction} 

Post-common-envelope binaries (PCEBs), which consist of a hot subdwarf B (sdB) star and a low-mass main-sequence companion in a close orbit, are important systems for studying binary evolution. The hot sdB components in these systems reside on the extreme horizontal branch of the Hertzsprung–Russell (H-R) diagram, burn helium in their cores, and retain only very thin hydrogen envelopes \citep{2009ARA&A..47..211H}. The tight orbital separations observed in these systems imply that they could not have maintained such configurations throughout their lifetimes, because the progenitor of the compact object would have engulfed its companion during the giant phase. Therefore, it is assumed that the system originated as a wider main-sequence binary with an orbital separation of about 1 AU. As the more massive component evolved into a giant, unstable mass transfer was triggered, leading to the formation of a common envelope (CE) surrounding both stars. The strong frictional drag forces within the CE cause the binary system to lose orbital energy and angular momentum. This leads the two stars to spiral inward. Once enough energy accumulated in the envelope, it was expelled from the system, leaving behind a binary consisting of the compact object and companion star \citep{1976IAUS...73...75P, 1984ApJ...277..355W, 2010A&A...520A..86Z, 2012ApJ...745L..23Q, 2013A&A...549A..95Z, 2016PASP..128h2001H, 2016MNRAS.459.4518H}.

Their short orbital periods and deep eclipses allow for the determination of eclipse timings with an accuracy of just a few seconds, enabling the detection of even minor variations in orbital periods. Such eclipse-timing variations (ETVs) have often been interpreted as evidence for the presence of a circumbinary planet \citep{2009ApJ...706L..96Q, 2010A&A...521L..60B, 2012ApJ...745L..23Q, 2012A&A...543A.138B, 2021MNRAS.507..809E, 2023MNRAS.526.4725O}. However, some ETV-based planetary architectures have been shown to be dynamically unstable or otherwise implausible \citep{2012MNRAS.427.2812H}, so such interpretations should be treated with caution. Circumbinary planets have been reported around several post-common-envelope sdB binaries. \citet{2013A&A...549A..95Z} examined orbital period variations in thirteen sdB systems and detected significant variations in five of them. Many researchers have interpreted these evidence for additional massive bodies such as those proposed for HS 0705+6700, HS 2231+2441, NSVS 14256825, NY Vir, and HW Vir \citep{2009ApJ...695L.163Q, 2010Ap&SS.329..113Q, 2012A&A...543A.138B, 2014MNRAS.445.2331L, 2017AJ....153..137N, 2021MNRAS.507..809E, 2024PASA...41...47E, 2025AdSpR..76.1204E}. \citet{2022MNRAS.514.5725P} re-investigated seven post-common-envelope binaries (four of which are sdB systems: HW Vir, NY Vir, HS 0705+6700, and NSVS 14256825) and demonstrated that the majority of the 
previously proposed models failed to align with the new \textit{O – C} diagrams, even over short intervals. More recently, \citet{2024PASA...41...47E} extended the \textit{O – C} baseline of HS 0705+6700 to twenty-four years and modeled its period variations using linear, quadratic, and additional-body terms. They found that, although models with multiple closely spaced brown dwarfs could statistically fit the data, these configurations would be unstable on short timescales. In this context, Kepler-451 serves as a well-suited test case for these interpretations, given its well-studied nature and extensive observational coverage.

The Kepler-451 system, also cataloged as 2MASS J1938+4603, KIC 9472174, TIC 271164763, 1SWASP J193832.60+460359.1, and NSVS 5629361, was first identified by \citet{2010MNRAS.408L..51O} who classified it as an sdB+dM post-common-envelope binary with an orbital period of $\sim$0.12576 days and component masses of $M_{1}=0.48 \pm0.03 \ M_{\odot}$ and $M_{2}=0.12 \pm0.01 \ M_{\odot}$. The orbital period variations of Kepler‐451 have been the subject of numerous studies. Using Kepler photometry, \citet{2012ApJ...753..101B} measured the Rømer delay, which causes eclipses to appear shifted depending on the orbital position of the star, and found that the secondary eclipse occurs about 2s behind the midpoint between the consecutive primary eclipses. \citet{2014A&A...566A.128L} analyzed the multi-year SuperWASP archive for twelve post-common-envelope binary systems, including Kepler-451. By comparing the mid-eclipse times with a previously published ephemeris, they found no clear evidence of significant period variation in the system. Using Kepler photometry, \citet{2015A&A...577A.146B} reported a periodic signal in the $O-C$ diagram, attributing it to a tertiary companion orbiting at 0.92 AU with a period of 416 days and a minimum mass of $\sim1.9 \ M_{Jup}$. They also identified a significant long-term trend in the timing data that could indicate an evolutionary effect or the presence of other companions. Both \citet{2015A&A...577A.146B} and \citet{2020A&A...642A.105K} derived mid-eclipse times from prewhitened Kepler light curves. The studies differ primarily in their datasets (SC-only vs. LC+SC) and timing analysis procedures. \citet{2020A&A...642A.105K} challenged the planetary interpretation by reanalyzing the Kepler-451 timing data with refined reduction techniques designed to mitigate the effects of sdB pulsations and to limit calibration/reduction systematics. In addition to the $O-C$ analysis, \citet{2020A&A...642A.105K} applied the binarogram method as an independent check. Neither approach recovered a significant modulation at the proposed $\sim$416-day period, thereby rejecting the Jupiter-mass companion suggested by \citet{2015A&A...577A.146B}. Subsequently, \citet{2020RNAAS...4..237B} reprocessed the timing data and confirmed a clear periodic ETV signal, supporting the initial detection. More recently, \citet{2022MNRAS.511.5207E} performed a long-baseline analysis that revealed more complex eclipse-timing behavior in Kepler-451. By combining Kepler eclipse timings with additional years of ground-based monitoring and TESS observations, \citet{2022MNRAS.511.5207E} detected multiple periodicities in the $O-C$ curve. The originally reported modulation of $\sim416$ days was refined to $\sim406$ days with a modest eccentricity. Two further signals were identified: an inner companion with a period of $\sim43$ days, present only in the Kepler residuals after a two-body model was fitted to the full $O-C$ diagram, and an outer body with a cycle of $\sim1800$ days. These variations were interpreted as evidence for a three-planet circumbinary system, with all three Jovian‐mass companions having comparable masses.

In this work, we extend the eclipse‐timing baseline of Kepler‐451 with our new measurements to investigate whether its period variations arise from circumbinary companions or alternative mechanisms, with the ultimate aim of offering alternative solutions to orbital configurations and understanding the dynamics of the system.

\section{Observations and Timing Data}
\label{sec:2}

We performed photometric observations of Kepler-451 from 2014 to 2024 using three ground-based telescopes. These data were obtained with the 1 m telescope equipped with a 4k $\times$ 4k SI1100 CCD with a pixel size of $15\times15$ $\mu$m at the Türkiye National Observatories (TUG100, Antalya, Türkiye), the 60 cm telescope equipped with an Andor iKon-M 934 CCD with a pixel size of $13\times13$ $\mu$m at the Adiyaman University Observatory (ADYU60, Adiyaman, Türkiye), and the 50 cm telescope equipped with a CMOS QHY268M Pro I camera with a pixel size of $3.76\times3.76$ $\mu$m at the Türkiye National Observatories (ATA050, Erzurum, Türkiye). All observations were conducted without filters (white light) with exposure times ranging from 5 to 30 s, depending on weather conditions and target brightness. Standard data reduction procedures, including bias, dark, and flat‐field corrections, were applied to all images, and differential aperture photometry was performed following \citet{2021MNRAS.507..809E}. Aperture photometry was performed with a Python script using a nearby reference star validated by a check star. Aperture radii from 1 to 3$\times$FWHM were tested, and 1.4$\times$FWHM was chosen for its stability and high S/N. In total, we obtained 68 primary eclipse light curves from our ground-based Kepler-451 campaign.

In addition to our ground-based observations, we used space-based observations from Transiting Exoplanet Survey Satellite (\textit{TESS}; \citep{2015JATIS...1a4003R}). \textit{TESS} observed Kepler-451 in sectors 14, 15, 40, 54, 74, 75, 81, and 82 from July 2019 to September 2024, providing 20- and/or 120-s cadence data depending on the sector. We used both cadences across different sectors. When both were available for the same eclipse, we ensured that no eclipse was counted twice by retaining only one measurement. Longer-cadence data (200, 600, or 1800 s) were not used. Following \citet{2020RNAAS...4..237B} and \citet{2022MNRAS.511.5207E}, we adopted the PDCSAP (Pre-search Data Conditioning) flux, which is corrected for common instrumental and long-term systematic trends using an optimized aperture. The temporal resolution of our data is determined by the integration time (20 s or 120 s), which is sufficient to resolve eclipses that would be undersampled in long-cadence (e.g., 1800 s) data. Using the Lightkurve package\footnote{\url{https://docs.lightkurve.org/}} \citep{2018ascl.soft12013L} to access the Barbara A. Mikulski Archive for Space Telescopes (MAST)\footnote{\url{https://mast.stsci.edu/}}, we retrieved PDCSAP flux time series, from which a total of 5095 primary eclipse light curves were extracted.

We derived the mid-eclipse times by fitting each eclipse light curve from our observations and \textit{TESS} data using the modified Gaussian function described in \citet{2012A&A...540A...8B}. The modified Gaussian model \citep{2012A&A...540A...8B} combines a Gaussian profile with a polynomial term to accurately fit eclipse light curves, reproducing the steep ingress and egress and the flat-bottom shape of the minima. The root‐mean‐square (RMS) residuals between the observed and modeled light curves ranged from 0.005 to 0.017 mag, with a mean of 0.008 mag. Furthermore, the largest and average errors of the mid-eclipse times obtained from our observations are 16.8 and 7.6 s, respectively, while the average error for the \textit{TESS} data derived in this study is 16.68 s. The mid‐eclipse times obtained from ground-based telescopes were then converted to barycentric dynamical Julian Date (BJD) using the procedure described by \citet{2010PASP..122..935E}.

\begin{table}
\caption{The Mid-eclipse Times of Kepler-451}
\label{tab:midtimes_kepler451}
\small
\centering
\begin{tabular}{lcl}
\hline
\multicolumn{3}{c}{TESS} \\
\hline\hline 
BJD & error &  References\\
\hline
2458683.4637516300 & 0.0002263355 & TESS\\
2458683.5897780000 & 0.0002893584 & TESS\\
2458683.7153597600 & 0.0002353802 & TESS\\
 ... & ... & ...\\
\hline
\multicolumn{3}{c}{From our data} \\
\hline\hline 
BJD & error &  References\\
\hline
2457016.1935801600 & 0.0001410417 & This work\\
2457017.1999030400 & 0.0000671462 & This work\\
2457106.6189058200 & 0.0000801022 & This work\\
... & ... & ...\\
\hline
\multicolumn{3}{p{.8\columnwidth}}{$^{*}$ The full table will be published in machine-readable format as online supplementary material.}\\

\end{tabular}
\end{table}

\section{LTT MODEL OF THE ORBITAL PERIOD VARIATION}

\begin{figure*}
\includegraphics[width=\textwidth]{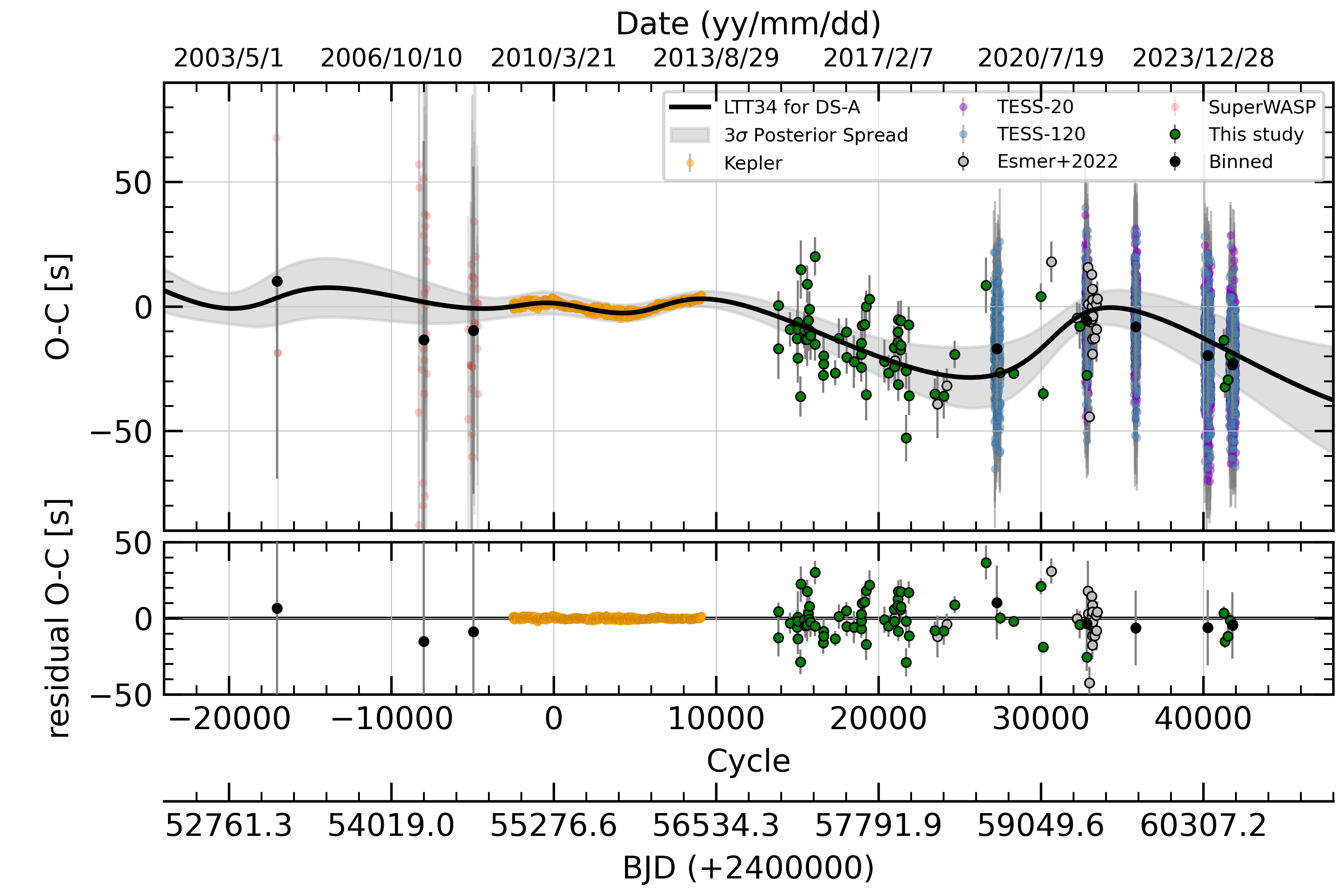}
\caption{The O–C diagram of Kepler-451 relative to the linear ephemeris from Equation \ref{eq:lineer_eph_DSA} for DS-A is shown in the upper panel. The solid black line represents the two-companion LTT model. The gray shaded area represents the $\pm3\sigma$ posterior spread calculated from 1000 randomly selected parameter samples from the MCMC posterior. The lower panel displays the timing residuals after subtraction of the full model, with an RMS scatter of 3.23 s.} 
\label{fig:OC_1}
\end{figure*}

\begin{figure*}
\includegraphics[width=\textwidth]{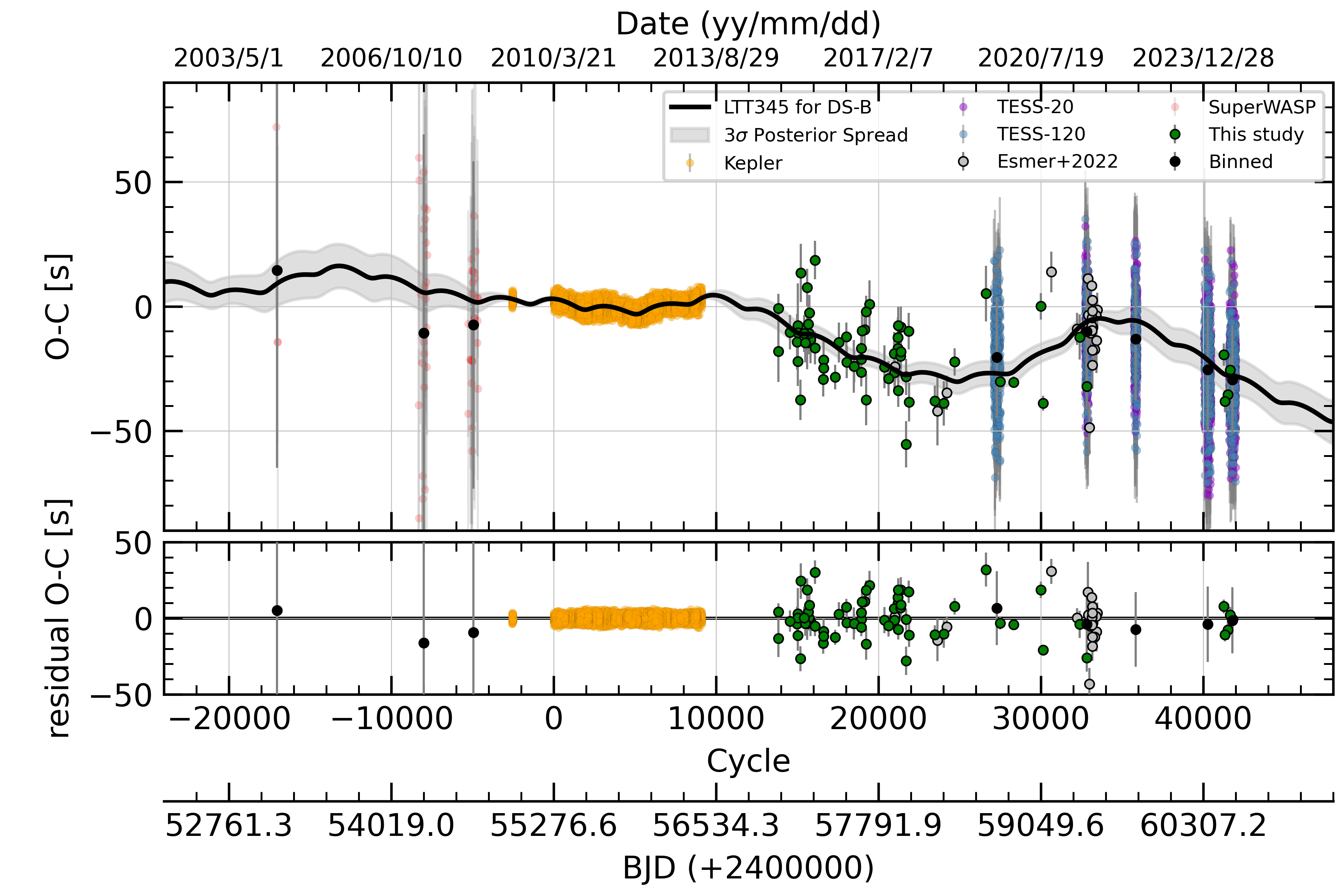}
\caption{The O–C diagram of Kepler-451 based on DS-B, fitted with the three-companion LTT model. The solid black line shows the model fit, while the lower panel shows the residuals after subtraction the complete model, with an RMS scatter of 2.31 s.} 
\label{fig:OC}
\end{figure*}

To investigate variations in the orbital period of Kepler-451, we constructed updated $O-C$ diagrams using two distinct data sets as shown in Fig. \ref{fig:OC_1} and \ref{fig:OC}, by combining our new mid-eclipse times with previously published values. Because secondary eclipses are significantly shallower than the primary eclipses, most studies focus on the primary mid-eclipse times. Following this convention, we used only the primary mid‐eclipse times in our analysis. The observational baseline was extended by incorporating our ground-based observations, SuperWASP eclipse times from \citet{2014A&A...566A.128L}, and previously unused TESS observations (sectors 54, 74, 75, 81, and 82). The resulting extended dataset significantly enhances the precision in detecting long-term period variations. We defined two distinct data sets: Data Set A (hereafter DS-A) consists of \textit{Kepler} data prewhitened for pulsations using the seven-period template fitting, provided by \citet{2020A&A...642A.105K}.

Data Set B (DS-B) includes the \textit{Kepler} mid-eclipse times derived by \citet{2022MNRAS.511.5207E} from short-cadence PDCSAP light curves by fitting a fixed light-curve model cycle by cycle, varying only the conjunction times and applying a brightness normalization parameter. Their approach did not include explicit prewhitening of pulsations prior to minima determination. These methodological differences motivated us to analyze them as separate data sets. The mid‐eclipse times from SuperWASP and TESS observations exhibit substantial scatter in the $O-C$ diagram. Therefore, we binned the data into cycle groups and used only these binned times in our modeling. The uncertainty of each binned value was estimated as the square root of the sum of the squares of the mean mid‐eclipse‐time error and the standard error. The complete list of mid-eclipse times from our ground-based and TESS observations is given in Table \ref{tab:midtimes_kepler451}. We then applied a weighted least-squares fit to derive the linear ephemeris parameters for DS-A and DS-B:

% \begin{equation}
% \label{eq:lineer_eph_DSA}
% \text{DS-A}: T_{eph}(E) = T_0 + L \times P_{bin}
%    =  {BJD}\: 2455276.608480 (23) 
%      + E \times 0.1257652725 (35) 
% \end{equation}

% \begin{equation}
% \label{eq:lineer_eph_DSB}
% \text{DS-B}: T_{eph}(E) = T_0 + L \times P_{bin}
%    =  {BJD}\: 2455276.608525 (24) 
%      + E \times 0.1257652722 (46) 
% \end{equation}

\begin{equation}
\label{eq:lineer_eph_DSA}
\begin{split}
\text{DS-A: } T_{eph}(E) &= T_0 + L \times P_{bin} \\
 &= \mathrm{BJD}\; 2455276.608480 (23) \\
 &\quad + E \times 0.1257652725 (35)
\end{split}
\end{equation}

\begin{equation}
\label{eq:lineer_eph_DSB}
\begin{split}
\text{DS-B: } T_{eph}(E) &= T_0 + L \times P_{bin} \\
 &= \mathrm{BJD}\; 2455276.608525 (24) \\
 &\quad + E \times 0.1257652722 (46)
\end{split}
\end{equation}

where $T_0$ is the initial ephemeris and $P_{bin}$ is the orbital period of the binary system for the mid-eclipse time at zero cycle ($L=0$). $T_{eph}$ is the mid-eclipse time in BJD. The resulting $O-C$ diagram of DS-A and B, based on their linear ephemeris, is presented in Figure \ref{fig:OC_1} and \ref{fig:OC}, respectively. 

We also tested a quadratic ephemeris to explore potential long-term trends in the orbital period. However, the resulting fits were not statistically satisfactory. Comparison of the coefficient of determination ($R^2$) between the linear and quadratic ephemeris fits shows that inclusion of the quadratic term significantly worsened the fit quality. For DS-A, the $R^2$ dropped from 0.9983 (linear) to 0.668 (quadratic). Similarly, for DS-B it decreased from 0.9999 (linear) to 0.9983 (quadratic). More importantly, the inclusion of a quadratic term distorted the overall shape of the $O-C$ diagram, imposing an artificial parabolic curvature that degraded the residual structure and misrepresented the periodic features evident in the data. Therefore, we did not adopt the quadratic ephemeris as a standalone solution. Nonetheless, a quadratic term ($\beta_{\text{eph}}$) was included as a free parameter in the full LTT model to account for any potential long-term trends, if statistically supported.

We tested various LTT configurations (two LTT terms, quadratic ephemeris with two LTT terms, three LTT terms, and quadratic ephemeris with three LTT terms) for both data sets. The most general form of the formulation is as follows:
\begin{equation}
 \label{eq:model}
 T_{eph}(L) = T_{0} + L \times P_{bin}  + \beta_{eph} \times L^2 + 
\sum_{i=0}^{i} \tau_{i}(L),
\end{equation}
where, $\beta_{eph}$ refers to a quadratic term which is computed as $P \dot{P}/2$. $\dot{P}$ is the derivative of the period over time (i.e., dP/dt).
$\tau_{i}$ term represents the LTT contribution of \textbf{$i^{th}$} companion, defined as:
\begin{equation}
\label{eq:tau}
 \tau_{i}= K_i \left(\sin{\omega_i} (\cos{E_i (t)} - e_i) + \sqrt{1-e_i^2} \cos{\omega_i} \sin{E_i (t)}\right),
\end{equation}
where $K_i$ is the semi-amplitude of the LTT signal of the \textbf{$i^{th}$} companion. 
The parameters $e_i$, $\omega_i$ and $t_{0,i}$ represent its eccentricity, the longitude of the pericenter, and the time of the pericenter passage, respectively, and $E_i$ denotes the eccentric anomaly. To avoid poorly constrained values of eccentricity ($e_i$) and argument of periastron ($\omega_i$) in nearly circular or moderately eccentric orbits, we used Poincaré elements defined as $x_i \equiv e_i \cos\omega_i$ and $y_i \equiv e_i \sin\omega_i$. For further details on these parameters, see \citet{2012MNRAS.425..930G, 2015MNRAS.448.1118G, 2017AJ....153..137N}.

We used Markov Chain Monte Carlo (MCMC) analysis to estimate the model parameters, following the approach of our previous studies \citep[e.g.][]{2021MNRAS.507..809E, 2023MNRAS.526.4725O}. Uniform priors were randomly assigned to each parameter within physically meaningful boundaries, such as $\beta, K_i, P_i, t_{0,i}, \sigma_f > 0$ days, $x_i,y_i \epsilon [-0.75,+0.75]$, $P_{bin},  \epsilon [0.08, 0.16]$ days, and $\Delta T_0 \epsilon [-10, +10]$ days. An additional parameter ($\sigma_f$) was included in the likelihood function to scale observational uncertainties.

The need for introducing the $\sigma_f$ parameter arose from the initially high reduced chi-square ($\chi^2_\nu \gg 1$, as reported in previous studies for this system), indicating that either the model did not fully capture all relevant timing variations or that the timing uncertainties were underestimated. The quasi-periodic nature of the $O-C$ variations (as in Fig.~\ref{fig:OC_1} -~\ref{fig:OC} ) and astrophysical considerations supported the plausibility of an LTT-based model. Therefore, to address the second possibility of underestimated uncertainties, we used a systematic uncertainty term analogous to the "stellar jitter" term used in radial velocity analyses \citep[e.g.][]{2005PASP..117..657W}, where $\sigma_f$ accounts for unmodeled astrophysical and instrumental noise sources. In our case, $\sigma_f$ was added in quadrature to the formal mid-eclipse timing uncertainties using the relation $\sigma_i^2 \rightarrow \sigma_i^2 + \sigma_f^2$, where $\sigma_i$ represents the timing uncertainty of the \textbf{$i^{th}$} measurement. This systematic term encompasses contributions from effects such as unmodeled eclipse geometry, variable atmospheric conditions, pulsation of a stellar companion, data reduction, or observational systematics. By treating $\sigma_f$ as a free parameter in the MCMC analysis, we allowed the model to self-consistently determine the level of additional uncertainty required to reconcile the fit with the data.  This correction brought the reduced chi-square close to unity and ensured statistically robust posterior distributions for the model parameters, following a similar approach as in \citet{2015MNRAS.448.1118G}.

The posterior distributions were sampled using the affine-invariant ensemble sampler provided by the \textit{emcee} package \citep{2010CAMCS...5...65G, 2013PASP..125..306F}. We used 512 walkers and ran the chains for up to 150,000 iterations, adjusting the complexity as required by the model.

\begin{figure*}
\includegraphics[width=\textwidth]{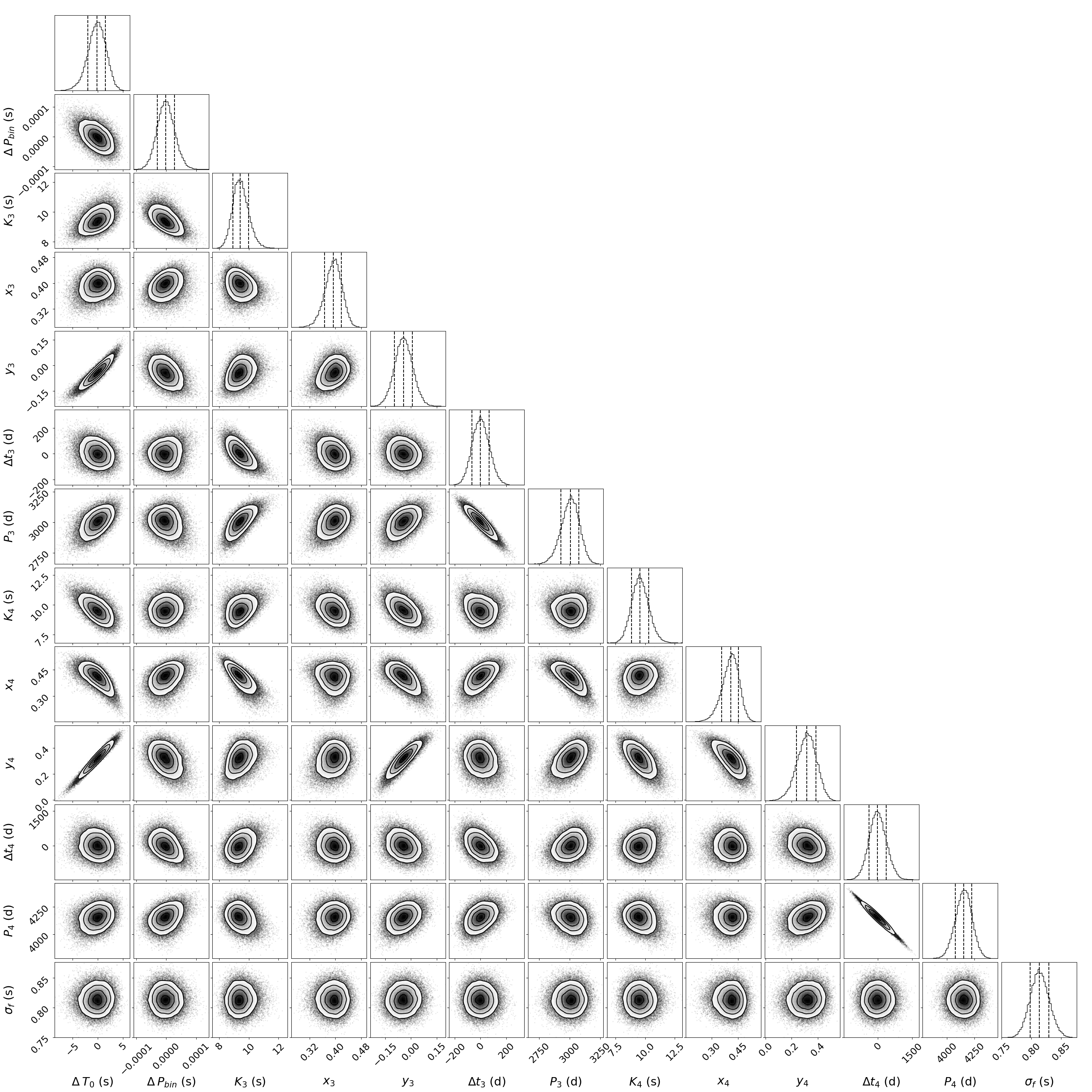}
\caption{The 1-D and 2-D projections of the posterior probability distributions of the free parameters extracted from the $O-C$ diagram for the model including two LTT terms for DS-A. This corner plot was generated with corner.py \citep{2016JOSS....1...24F}. } 
\label{fig:corner_1}
\end{figure*}

\begin{figure*}
\includegraphics[width=\textwidth]{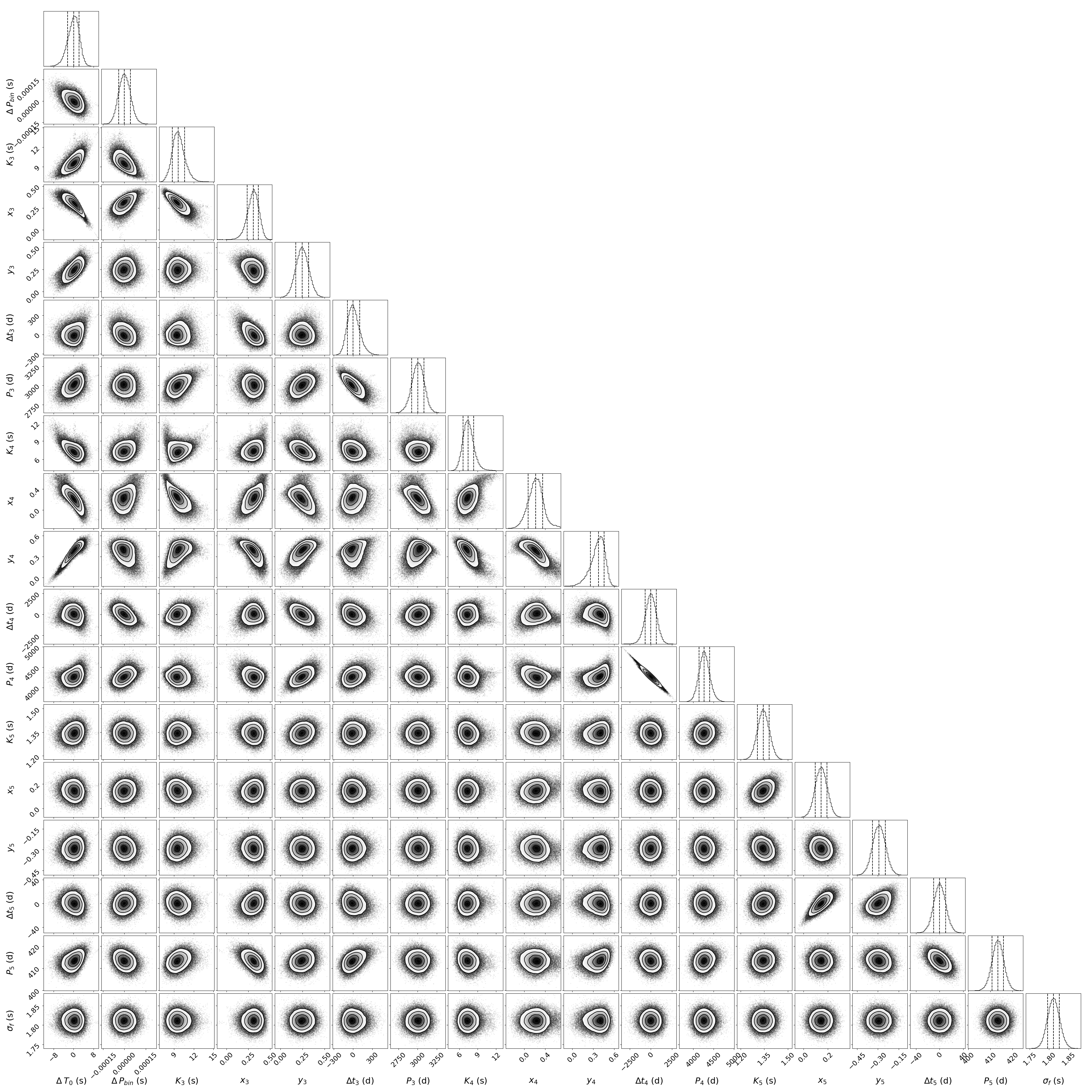}
\caption{The 1-D and 2-D projections of the posterior probability distributions of the model including three LTT terms for DS-B. } 
\label{fig:corner_2}
\end{figure*}

Based on statistical results, we propose that the two-companion LTT model (LTT34) provides the most reasonable solution for DS-A, whereas the three-companion LTT model (LTT345) provides the best fit for DS-B. For DS-A, the LTT34 model yielded an RMS scatter of 3.23 s and a reduced chi-square ($\chi^2_\nu$) of 1.23. Introducing the third companion (LTT345 model) did not produce a significant improvement (RMS=3.24), and did not justify the added complexity compared to the simpler two-companion model (LTT34), thus rendering the additional complexity unnecessary. In contrast, for DS‐B, the LTT345 model achieved better results, with an RMS of 2.31 s and a $\chi^2_\nu$ of 1.01. Although we also tested the LTT34 model for DS-B, the MCMC chains failed to converge to a well-defined solution. The resulting posterior distributions were multi-modal and poorly constrained, indicating that the model did not yield a statistically meaningful or physically interpretable fit. Therefore, we do not present the results for this model. The final model parameters and their uncertainties, derived from the 16th, 50th, and 84th percentiles of the posterior distributions, are given in Table \ref{tab:results}. Figure \ref{fig:corner_1} and \ref{fig:corner_2} present the corner plots, showing the one- and two-dimensional posterior probability distributions of the fitted parameters, which are well-constrained around single, normally distributed solutions. The MCMC analysis converged to $\sigma_f = 0.81$ s for DS-A and $\sigma_f = 1.81$ s for DS-B, quantifying the level of additional uncertainty required to account for unmodeled noise/systematic sources.

\begin{table*}
\caption{System Parameters for Kepler-451. }
\label{tab:results}
\begin{tabular}{lllll}
\hline\hline 
& &  & DS-A & \\
\hline 
Parameters & Unit &  Inner Companion (3rd body) & Middle Companion (4th body) & Outer Companion (5th body)\\
\hline
$a_isin_i$         & au       & $3.441^{+0.218}_{-0.174}$ & $4.268^{+0.383}_{-0.318}$ &   \\

$e_i$             &           & $0.398^{+0.024}_{-0.026}$ & $0.522^{+0.034}_{-0.028}$   &    \\

$\omega_i$        & deg       & $-5.59^{7.22}_{-8.28}$ & $38.29^{9.24}_{-8.76}$ &  \\

$t_{0,i}$         & BJD       & $2453083^{+72}_{-71}$  & $2434353^{+410}_{-363}$ &  \\

$P_i$             & yr        & $8.24^{+0.21}_{-0.22}$ &  $11.38^{+0.22}_{-0.23}$ &  \\

$K_i$             & s         & $9.42^{+0.63}_{-0.48}$ & $9.52^{+0.81}_{-0.65}$ &  \\

$M_i\sin{i_i}$    & $M_{Jup}$ & $3.75^{+0.83}_{-0.84}$ &  $3.081^{+0.83}_{-0.82}$ &  \\
\hline\hline
& &  & DS-B & \\
\hline 
Parameters & Unit &  Inner Companion (3rd body) & Middle Companion (4th body) & Outer Companion (5th body)\\
\hline
$a_isin_i$      & au        & $0.916^{+0.029}_{-0.028}$    & $3.437^{+0.30}_{-0.25}$    & $4.344^{+0.635}_{-0.522}$     \\
$e_i$           &           & $0.326^{+0.047}_{-0.046}$    & $0.391^{+0.059}_{-0.055}$    & $0.434^{+0.112}_{-0.111}$     \\
$\omega_i$      & deg       & $-63.693^{+4.202}_{-4.201}$  & $38.522^{+12.810}_{-12.812}$ & $59.275^{+10.110}_{-10.109}$  \\
$t_{0,i}$       & BJD       & $2455114^{+10}_{-11}$        & $2453373^{+100}_{-86}$       & $2433841^{+643}_{-660}$       \\
$P_i$           & yr        & $1.13^{+0.006}_{-.0.007}$    & $8.23^{+0.21}_{-0.22}$       & $11.69^{+0.36}_{-0.35}$       \\
$K_i$           & s         & $1.34^{+0.04}_{-0.03}$       & $9.60^{+0.98}_{-0.86}$       & $7.52^{+0.92}_{-0.82}$        \\
$M_i\sin{i_i}$  & $M_{Jup}$ & $1.86^{+0.06}_{-0.06}$       & $3.70^{+1.18}_{-1.17}$       & $2.20^{+0.82}_{-0.83}$        \\
\hline\hline
\end{tabular}
\end{table*}

\section{Magnetic Mechanism and Orbit Stability}

To test whether the observed timing amplitudes in our O–C solutions for Kepler-451 could be produced by non-planetary mechanisms, we investigated the Applegate mechanism, in which magnetic activity cycles in the secondary star lead to cyclic variations in the orbital period \citep{1992ApJ...385..621A, 1998MNRAS.296..893L}.  The mechanism proposed by \citet{1992ApJ...385..621A} suggests that solar-like magnetic cycles can redistribute angular momentum within the star, altering its quadrupole moment and inducing periodic orbital period changes. \citet{2009Ap&SS.319..119T} reformulated this approach (hereafter Standard model) by integrating the original concept with the improved model of \citet{1998MNRAS.296..893L}, which accounts for variations in the stellar quadrupole moment caused by both rotational and magnetic energy.  \citet{2016A&A...587A..34V} developed the “Two-zone” model, incorporating distinct density profiles for the stellar core and shell, and applying the method to realistic stellar structures. This model typically predicts a much higher energy requirement to drive the mechanism.  More recently, \citet{2020MNRAS.491.1820L} proposed an alternative formulation (hereafter Spin-orbit coupling model), in which the modulation of the orbital period arises from the coupling between the stellar spin and the orbital motion, driven primarily by changes in the non-axisymmetric quadrupole moment rather than by long-term tidal effects. This approach significantly reduces the required energy compared to earlier models.

For each Applegate-like model (Standard, Two-zone, and Spin-orbit coupling), we calculated the required energy, $\Delta E$, as a fraction of the total energy available in the magnetically active secondary star, $E_{sec}$, using the system parameters given in Table \ref{tab:results} and the stellar parameters of Kepler-451 (e.g. $M_2=0.12 \ M_\odot$ and $R_2=0.17 \ R_\odot$) reported by \citet{2022MNRAS.511.5207E}. In all Applegate-like models, a ratio $\Delta E/E_{sec} < 1$ indicates that the magnetic mechanism could be energetically sustainable. Ratios close to or slightly exceeding unity imply that the mechanism remains uncertain but cannot be entirely ruled out, whereas values $\gg 1$ effectively exclude a magnetic origin. The results of the Applegate mechanism analysis for each LTT term are summarized in Table \ref{tab:applegate}. In the Two-zone and Spin–orbit coupling models, all LTT terms in both DS-A and DS-B require energy budgets that substantially exceed the available energy, effectively ruling out a magnetic origin. The energy budget implied by the LTT signals of the fourth companion in DS-A and the fifth companion in DS-B results in $\Delta E/E_{\rm sec}<1$ under the \emph{Standard} model, indicating that a magnetic contribution cannot be excluded for these two cases.

\begin{table}[ht!]
\caption{The energy ratios for the formulation of corresponding magnetic mechanisms \citep{2009Ap&SS.319..119T,2016A&A...587A..34V,2020MNRAS.491.1820L}.}
\label{tab:applegate}
\centering
\small
\begin{tabular}{lccc}

\hline\hline 
LTT Signal & \multicolumn{3}{c}{$\Delta E$/$E_{sec}$}  \\
  & Standard & Two-zone & \shortstack{Spin-orbit\\coupling} \\
\hline
4th Body (for DS-A)  & 0.33 & 5.94 & 15.11 \\
5th Body (for DS-B)  & 0.18 & 3.31 & 11.15 \\
\hline
\multicolumn{4}{p{.8\linewidth}}{$^{*}$ We only included the LTT signals with the minimum $\Delta E$/$E_{sec}$.}\\
\end{tabular}

\end{table}

To investigate the long-term dynamical orbital stability of the system, we used the N-body orbital integration package of REBOUND\footnote{https://rebound.readthedocs.io} \citep{2012A&A...537A.128R}, which simulates the motion of celestial bodies. It provides two important outcomes: the orbital stability timeline (OST) with the WHFast integrator, and the surface map with the MEGNO chaos indicator (hereafter Megno map). The WHFast indicator uses a Wisdom–Holman symplectic method that conserves energy exceptionally well by splitting each timestep into Keplerian and interaction phases \citep{2015MNRAS.452..376R}. The OST shows how parameters such as the semimajor axis and eccentricity change over time by integrating the orbits over a given period. The MEGNO indicator ($\langle Y \rangle$) provides the average exponential divergence of nearby orbits and assigns a numerical value that definitively separates regular motion from chaotic behavior ($\langle Y \rangle > 2$) \citep{2000A&AS..147..205C, 2015MNRAS.452..376R}. The MEGNO surface map reveals regions of chaotic behavior within the orbital configuration. Additionally, the OST timeline offers an accurate, computationally efficient representation of the interactions and long-term dynamical evolution of these multi-body systems (For more detail see \citet{2001A&A...378..569G, 2021MNRAS.506.2122B, 2023MNRAS.526.4725O, 2025NewA..11902414E}).

Studies of the stability of circumbinary planets often assume that all planetary orbits lie in the orbital plane of the binary. This approach is supported by both theory and observation. The classic stability criteria of \citet{1999AJ....117..621H} are based on the assumption of coplanar orbits. Moreover, circumbinary planets identified by Kepler have inclinations of only a few degrees relative to their host binaries \citep{2019A&A...624A..68M, 2019A&A...628A.119C}. Accordingly, we treated the binary system as a single central mass and assumed all orbits to be coplanar.

We ran the integration over $10$ Myr using the best-fit orbital parameters derived from both data sets (Table \ref{tab:results}).  For DS-A, we adopted the parameters from the two-companion LTT model (LTT34). 
Strong mutual gravitational perturbations between the companions led to dynamical instability within a few thousand years, with large oscillations in semi‐major axes and eccentricities. MEGNO values exceeded 2, indicating chaotic orbits.
Similarly, we evaluated the dynamical stability of the three-companion LTT model (LTT345) obtained from the DS-B data set. Gravitational interactions among the three additional bodies led to chaotic dynamical behavior. The orbital configuration of the system becomes disrupted within just 5000 years, primarily due to the early ejection of the inner companion, accompanied by chaotic variations in the semi-major axes and eccentricities of the companions. We also tested the dynamical stability of the three-companion model using the parameters of \citet{2022MNRAS.511.5207E}, integrating over a timescale of $10$ Myr. The system became highly unstable within a short timescale of $10^4$ years, and the MEGNO values remain above two, indicating chaotic behavior. \citet{2022MNRAS.511.5207E} found that their best-fit three-companion solution for the circumbinary system Kepler-451 was dynamically unstable due to strong mutual perturbations among the three planets. In N-body integrations of the best-fit orbits, the orbital elements of the planets exhibited chaotic oscillations, indicating an unstable configuration. The likely cause of this instability was the proximity of the outer planet. \citet{2022MNRAS.511.5207E} suggested that the three-planet system would be stable only if the orbit of the outer planet were expanded by about 0.3 AU, thereby avoiding the high-eccentricity interactions that caused the instability in the more compact configuration.

\begin{figure*}
\label{fig:stability}
\includegraphics[width=\textwidth]{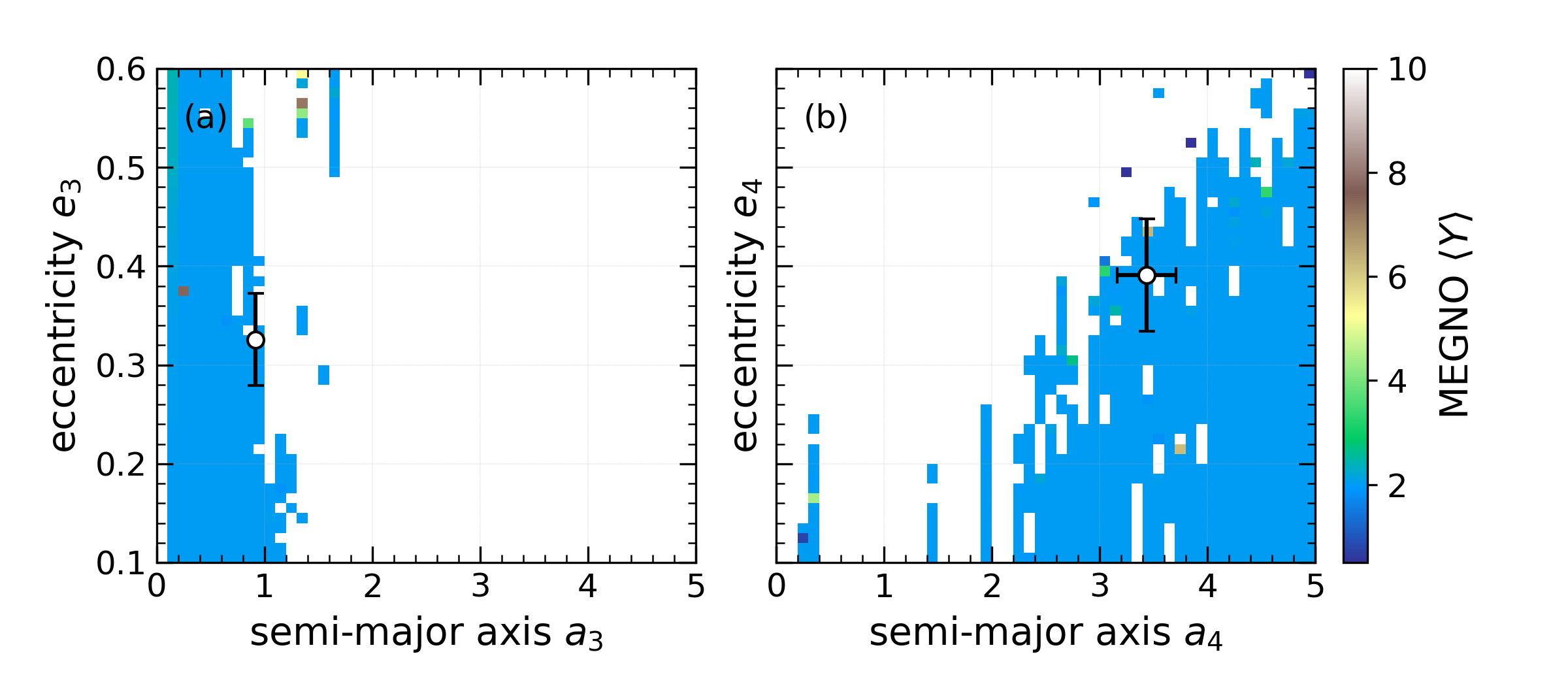}
\caption{Map of the MEGNO chaos indicator for the two-companion configuration for $10^7$ yr obtained using the system parameters for DS-B in Table \ref{tab:results}. The MEGNO chaos parameter is consistent with a borderline stable region.} 
\end{figure*}

The signals attributed to the forth body in the LTT34 of DS-A and the fifth body in the LTT345 of DS‐B may arise from magnetic activity. When the fourth-body signal in DS-A is excluded, the resulting configuration of the eclipsing binary with a single circumbinary companion remains dynamically stable for at least $10^7$ years. Given the relatively simple architecture of the system (2 + 1), such long-term stability is expected, and therefore no detailed stability maps are provided for this case. In order to assess the long-term dynamical behavior of the remaining companions in DS-B, we considered a two-companion system affected by the magnetic activity of the second star and performed an orbital stability analysis for this configuration. The $N$‐body simulations indicate that the resulting quadruple system (2 + 1 + 1) remains stable for at least $10^7$ years according to the orbital stability timeline. Furthermore, the MEGNO stability map (Figure \ref{fig:stability}) shows that the orbital solutions for the third body lie within the stable region ($\langle Y \rangle\leq2$), while those for the fourth body are located near the stability boundary. However, considering uncertainties, both companions could still be consistent with orbital stability.

\section{Discussion and Conclusions}
Using an extended observational baseline from 2004 to 2024 that combines mid‐eclipse timings from both ground‐ and space‐based observations, we investigated the orbital period variations of Kepler‐451 through detailed LTT modeling. Two independent datasets (DS-A and DS-B) were analyzed to account for differences in eclipse timing extraction methods from the Kepler light curves. Across both datasets, a variety of LTT configurations were tested, revealing that the preferred model depended on the choice of dataset. For DS-A, the two-companion configuration (LTT34) provided a statistically consistent explanation of the observed $O-C$ variations, while for DS-B, a three-companion configuration (LTT345) yielded the best overall statistical fit. The differing outcomes highlight that the adopted light-curve analysis method and associated systematics can influence the inferred system architecture.

For DS-A, adding a fifth body to form the LTT345 configuration did not yield a meaningful improvement in fit quality over LTT34, with RMS changing from 3.23 s to 3.24 s and reduced $\chi^2_\nu$ values from 1.23 to 1.28, remaining essentially unchanged. Moreover, the systematic errors for the LTT34 and LTT345 models were similarly small (0.81 s and 0.78 s, respectively). In DS-B, the LTT345 model with $\chi^2_\nu = 1.01$ and RMS = 2.31 s performed better than simpler models and also improved upon the RMS of 2.63 s obtained when applying the model of \citet{2022MNRAS.511.5207E} to our extended dataset. However, the relatively large systematic error (1.81 s) in DS-B, combined with the low semi-amplitude of the third body of  1.34 s, which was smaller than the systematic error itself, suggested that this signal could also be attributable to observational and calibration systematics rather than a genuine dynamical companion. Given the discrepancy between DS-A and DS-B, the periodic signal of $\sim400$-day may have been the result of data reduction and calibration, particularly for high-timing resolution data from Kepler.
\citet{2020A&A...642A.105K} also implied that differences in data reduction and calibration can bias long-period trends. In particular, $\sim400$-day signal reported by \citet{2015A&A...577A.146B} disappeared once pulsations were prewhitened. We therefore regard this periodicity as uncertain and do not interpret it as strong planetary evidence. According to \citet{2021A&A...650A.205V}, single sdB stars exhibit a low incidence of planetary companions, with no confirmed detections to date, and space-based surveys are beginning to set upper limits. By contrast, ETV studies of sdB+dM binaries such as HS 0705+6700, HS 2231+2441 and NSVS 14256825 have yielded several candidate circumbinary companions, though many remain unconfirmed and may be affected by data reduction systematics.

In DS-A, the two-companion LTT model (LTT34) yielded orbits of moderate eccentricity with $e_3 = 0.398$ and $e_4 = 0.522$, while in DS-B, the three-companion model (LTT345) produced $e_3 = 0.326$, $e_4 = 0.391$, and $e_5 = 0.434$, values comparable to those reported in other circumbinary systems \citep{2012A&A...540A...8B, 2021MNRAS.506.2122B, 2015MNRAS.448.1118G}. Although the orbital parameters of our innermost planet are similar to those of the middle planet proposed by \citet{2022MNRAS.511.5207E}, our analysis indicates two outer companions with orbital periods of $\sim8.23$ and $\sim11.69$ years, substantially longer than previously reported, suggesting a wider orbital configuration. An important distinction from earlier works \citep{2012ApJ...753..101B, 2015A&A...577A.146B, 2022MNRAS.511.5207E} is that the observed timing variations were reproduced by distinct periodic signals without invoking a quadratic ephemeris term.

To assess whether any of the observed LTT signals can be caused by magnetic activity cycles in the secondary star rather than planetary companions, we applied three formulations of the Applegate mechanism.
The Two-zone and Spin-orbit coupling formulations yielded energy ratios $\Delta E / E_{\text{sec}}$ well above unity for all LTT terms, suggesting that most signals are unlikely to be magnetic in origin.
However, the Standard model formulation produced $\Delta E / E_{\text{sec}}<<1$ ratios for the fourth body in DS-A ($0.33$) and the fifth body in DS-B ($0.18$), both of which had similar amplitudes and periods.
These results indicated that while a magnetic origin was disfavored for Two-zone and Spin-orbit coupling models, it can not be entirely excluded for the Standard model cases. It should be noted that this discrepancy reflects the different assumptions underlying the Applegate mechanisms.
\citet{2018A&A...611A..48P} showed that M dwarfs exhibit multi-year magnetic cycles capable of modulating orbital periods. In a broader survey, \citet{2022MNRAS.514.5725P} concluded that only long-period, low-amplitude signals fit within the Applegate energy budget, while larger ratios require circumbinary companions. For example, direct imaging of V471 Tau ruled out a third body to explain its $\sim25$ yr modulation \citep{2015ApJ...800L..24H}, leaving magnetic modulation as the only viable explanation \citep{2018A&A...615A..81N}. 
If these signals in our study are indeed magnetic in nature, they would have provided a plausible pathway to reconcile the observed variations with a dynamically stable planetary configuration.

Dynamical stability simulations further constrained the viable configurations. Our simulations predicted rapid instability for the two-companion configuration of DS-A and the three-companion configuration of DS-B, caused by the compact spacing of the orbits and resulting mutual perturbations. This is consistent with the predictions of standard Hill sphere criteria \citep{1879MNRAS..39..258H, 1993Icar..106..247G, 1999AJ....117..621H, 2015ApJ...807...44P}. When the LTT terms of the outer companions in DS-A and DS-B are attributed to magnetic activity, the remaining configurations simplify to stable architectures: in DS-A, a binary with one circumbinary companion at 3.441 AU ($\sim$3.75 $M_{Jup}$), and in DS-B, a binary with two companions at 0.916 AU ($\sim$1.9 $M_{Jup}$) and 3.437 AU ($\sim$3.7 $M_{Jup}$). In both cases, the systems remain dynamically stable for at least $10^7$ years.
These hybrid scenarios, in which the outer LTT terms are attributed to magnetic modulation while the remaining circumbinary companions are treated as gravitationally bound planets, offer consistent explanations for both DS-A and DS-B that satisfy the statistical fits as well as the dynamical stability constraints.

The observed variations in the orbital periods of post-common-envelope binaries (PCEBs) may originate from either planets that have survived the violent envelope ejection phase (first-generation planets) or from planets formed subsequently in the remnant gas disk (second-generation planets). In our multi-companion model of Kepler-451, both companions have masses below $\sim6 M_{Jup}$, which satisfies a necessary, but not sufficient, condition for survival during the common-envelope (CE) phase \citep{2010MNRAS.408..631N, 2013MNRAS.432..500N}. Therefore, it is highly unlikely that these planets represent first-generation survivors. Indeed, the direct survival of Jovian-mass planets during CE evolution typically requires greater orbital energy than such companions can realistically supply, rendering a first-generation origin both energetically and dynamically implausible. Consequently, ETVs observed in PCEBs are better explained by either second-generation planets that formed within residual gas and dust disks or by magnetic activity cycles (Applegate-type mechanism) causing cyclic variations in the quadrupole moment of the convective secondary star. This interpretation was further supported by previous studies of systems such as NN Ser and HW Vir, where multi‐year eclipse timing signals and dynamical stability analyses favored second‐generation planet formation over first‐generation survival \citep{2013A&A...549A..95Z, 2016RSOS....350571V}. For example, \citet{2016MNRAS.459.4518H} identified convincing observational evidence of dust around NN Ser, supporting the existence of a circumbinary disk, a necessary condition for second‐generation planet formation. Additionally, \citet{2009AJ....137.3181L} demonstrated that second-generation circumbinary planets can remain dynamically stable over long timescales, though dynamical simulations of \citet{2012MNRAS.427.2812H} showed that specific orbital architectures, such as those proposed for HW Vir, can become unstable on short timescales. In the case of Kepler-451, the masses and orbital radii of our circumbinary companions are consistent with the expected characteristics of planets formed within a circumbinary disk. The longer‐period signals in both models of DS-A and DS-B are unlikely to be planetary in origin, as the associated energy requirement is well within the range explained by cyclic magnetic activity. Thus, it can be concluded that Kepler-451 hosts second-generation circumbinary planet(s) accompanied by an additional modulation driven by the cyclic magnetic activity of the M-dwarf secondary. Although not definitive proof of a planetary origin, this interpretation supports the second-generation planet formation scenario involving material ejected during the CE phase.

Given the orbital semi-major axes obtained from our current LTT models for the proposed companions, their projected angular separations at the distance of Kepler-451  \citep[$\approx$400 pc,][]{2012arXiv1211.7121R} were estimated to be only about 2.3-10.9 mas \citep{2005A&A...436..373P}. These values are several times smaller than the diffraction limits of current large ground-based telescopes equipped with extreme adaptive optics instruments, such as $\sim$40 mas for the adaptive optics system at the Keck II Observatory \citep{2000PASP..112..315W}, $\sim$55 mas in the $K$ band for SPHERE at the Very Large Telescope \citep{2019A&A...631A.155B}, $\sim$55 mas for  AO188+IRCS at the Subaru Telescope \citep{2010SPIE.7736E..3NM}, and $\sim$60 mas for the Gemini Planet Imager at Gemini South Telescope \citep{2014PNAS..11112661M}. Projected separations of this nature are significantly below the current resolving power of direct imaging facilities. Consequently, as indicated by the orbital parameters derived from the existing LTT models, confirmation of these companions through direct imaging remains unfeasible at present.

Due to the discrepancy between the results from the two databases, the observed periodic modulation for the inner companion in the model of DS-B could be attributable to observational and calibration systematics, whereas the longer-period modulation (i.e., corresponding to the outer companion) in both DS-A and -B models may have a magnetic origin. Moreover, simplified orbital configurations indicate long-term dynamical stability. Taken together, the most consistent interpretation of the current data provides strong evidence for a circumbinary planetary companion at $\sim$3.4 AU in Kepler-451, which likely formed after the common-envelope phase. The origin of the other observed LTT signals remains uncertain. Importantly, our dynamical analysis shows that instability arises only when all signals are interpreted as additional companions. If the longest signal is attributed to magnetic activity, the remaining configuration remains stable.
Future timing, radial velocity, and -where feasible- direct imaging observations will provide crucial independent tests of the planetary interpretation, helping to refine orbital parameters and to discriminate between planetary and non-planetary origins.

\section*{Acknowledgements}

This work includes data obtained through the Scientific and Technological Research Council of Turkey (TUBITAK), through project number 114F460 (I.N., H.E.). In this study, observational data obtained within the scope of the project numbered T100-631 and T100-1333, conducted using the TUG100 (TÜBİTAK National Observatory, Antalya) and ATA50 telescopes (Türkiye National Observatories, Erzurum) at the site under the Türkiye National Observatories, have been utilized, and we express our gratitude for the invaluable support provided by the Türkiye National Observatories, the observation team and all its staff. Funding for the ATA50 telescope has been provided by Atatürk University (P.No. BAP-2010/40). We also wish to thank Adiyaman University Astrophysics Application and Research Center (Türkiye) for the allocation of their telescope time. This paper includes some of the data collected by the \textit{TESS} mission, which are publicly available from the Barbara A. Mikulski Archive for Space Telescopes (MAST) operated by the Space Telescope Science Institute (STScI). Funding for the \textit{TESS} mission is provided by the NASA Science Mission Directorate. This work was supported by the BAGEP Award of the Science Academy. We would also like to thank Dr. Jerzy Krzesiński and Dr. Adam Blokesz for answering our questions about their Kepler data.

\section*{Data Availability Statement}
The data underlying this article are available in the article
and in its online supplementary material. Some of the data presented in this paper were obtained from the Mikulski Archive for Space Telescopes (MAST) at the Space Telescope Science Institute. The specific observations analyzed can be accessed via \dataset[https://doi.org/10.17909/62bc-b020]{https://doi.org/10.17909/62bc-b020}. STScI is operated by the Association of Universities for Research in Astronomy, Inc., under NASA contract NAS5–26555. Support to MAST for these data is provided by the NASA Office of Space Science via grant NAG5–7584 and by other grants and contracts.

\bibliography{sample701}{}
\bibliographystyle{aasjournalv7}



\end{document}